\documentclass[%
aps,
prx,
reprint,
superscriptaddress,
nofootinbib,
amsmath,
amssymb
]{revtex4-2}
\usepackage{
    physics,
    graphicx,
    multirow,
    tabularx,
    mathtools, 
    placeins, 
    nicefrac, 
    hyperref, 
    threeparttable, 
    siunitx,
    enumitem,
    makecell
}

\usepackage{mathtools}

\usepackage[capitalize]{cleveref}
\Crefname{section}{Sec.}{Secs.}

\usepackage[dvipsnames]{xcolor}
\hypersetup{
    colorlinks,
    linkcolor={blue!50!black},
    citecolor={blue!50!black},
    urlcolor={blue!80!black}
}

\usepackage[caption=false]{subfig} 
\begin{document}

\title{Erasure conversion for singlet-triplet spin qubits enables\\ high-performance shuttling-based quantum error correction}

\newcommand{\qmaddress}{\affiliation{Quantum Motion, 9 Sterling Way, London N7 9HJ, United Kingdom}}
\newcommand{\oxddress}{\affiliation{Department of Materials, University of Oxford, Parks Road, Oxford OX1 3PH, United Kingdom}}
\newcommand{\uclddress}{\affiliation{Department of Physics and Astronomy, University College London, Gower St, London WC1E 6BT, United Kingdom}}

\newcommand{\comHamza}[1]{{\color{brown} #1}}

\author{Adam Siegel}
\email{adam@quantummotion.tech}
\qmaddress
\oxddress

\author{Simon Benjamin}
\qmaddress
\oxddress

\date{\today}

\begin{abstract}
Fast and high fidelity shuttling of spin qubits has been demonstrated in semiconductor quantum dot devices. Several architectures based on shuttling have been proposed; it has been suggested that singlet–triplet (dual-spin) qubits could be optimal for the highest shuttling fidelities. Here we present a fault-tolerant framework for quantum error correction based on such dual-spin qubits, establishing them as a natural realisation of erasure qubits within semiconductor architectures. We introduce a hardware-efficient leakage-detection protocol that automatically projects leaked qubits back onto the computational subspace, without the need for measurement feedback or increased classical control overheads. When combined with the XZZX surface code and leakage-aware decoding, we demonstrate a twofold increase in the error correction threshold and achieve orders-of-magnitude reductions in logical error rates. This establishes the singlet–triplet encoding as a practical route toward high-fidelity shuttling and erasure-based, fault-tolerant quantum computation in semiconductor devices.
\end{abstract}

\maketitle

\section{Introduction}

Quantum error correction (QEC) is one of the cornerstones of practical quantum computing. Using logical objects constituted of a greater number of physical qubits, QEC promises to reduce the logical error rate to low enough levels for the execution of deep quantum algorithms. More specifically, if the rate of failure of the physical components is below a given threshold, scaling up an error correcting code such as the surface code guarantees a highly desirable exponential suppression of the errors.
Generally QEC is implemented by measuring parity operators, or so-called \textit{stabilisers}. While simple in appearance, each measurement circuit may be repeated many billions of times during the course of a computation. It is therefore paramount to build a quantum computer upon a technology that supports a fast implementation of such stabilisers, with high fidelity so as to be below the code's threshold, and in a fashion that is fully scalable.

Semiconductor spin qubits are an excellent candidate for such tasks thanks to their long coherence times~\cite{Struck_2020}, scalability~\cite{gonzalez-zalbaScalingSiliconbasedQuantum2021}, and compatibility with both advanced manufacturing technologies~\cite{Maurand_2016, Zwerver_2022, swift_large_scale_2025, swift_superinductor_2025, clarke_readout_2025} and cryogenic classical electronics~\cite{XueXiao2021Ccco, ruffino2021integrated, thomas_rapid_2025}. Compact silicon spin qubit processors have already been developed~\cite{Philips_2022}, and small-scale error correction codes have been realised~\cite{Takeda_2022}. High-fidelity single- and two-qubit gates exceeding $99\%$ have been demonstrated~\cite{Xue_2022, Noiri_2022, Mills_2022}, including in commercially fabricated chips~\cite{steinacker2024300mmfoundrysilicon}, with fidelities surpassing $99.9\%$ also reported~\cite{yonedaQuantumdotSpinQubit2018}. Among the available two-qubit operations, the CZ and CNOT gates stand out as particularly well-suited for stabiliser measurements. Remarkably both these operations (among others) are well-established as primitives in silicon, without requiring e.g. basis change. 

Additionally, spin qubits have demonstrated exceptional shuttling capabilities, with recent experiments succeeding in shuttling electrons across a silicon spin qubit structure with $99.99\%$ fidelity per shuttling increment and at a high speed~\cite{desmet2024highfidelitysinglespinshuttlingsilicon,binder_shuttling_2025}. This is backed by theoretical works that promise that even higher fidelities could be achieved~\cite{Langrock_2023}. Shuttling has been shown to provide notable benefits, such as mitigating dephasing noise~\cite{burkardSemiconductorSpinQubits2021}, enabling motional averaging, and aligning qubit frequencies to facilitate global gate operations~\cite{jnane2025harnessingelectronmotionglobal}. This potential can therefore have profound consequences for semiconductor-based quantum computers, as it enables longer-range connectivities, which can in turn be used to reduce compilation overheads or even for the construction of high-rate error correcting codes. 

Architectures for fault-tolerant computing have been proposed that extensively exploit shuttling, replacing the more conventional picture of a static quantum computer to one in which qubits are constantly moving. Architectures such as the Crossbar~\cite{li_crossbar_2018}, Spiderweb~\cite{boterSpiderwebArraySparse2022}, and the SpinBus~\cite{kunne_spinbus_2024} paradigms propose to shuttle ancilla and/or data qubits to effectively extend interactions ranges, creating more space for control electronics. Typically, operations such as initialisation, measurement and gates are performed in manipulations zones between which qubits are shuttled~\cite{boterSpiderwebArraySparse2022, li_crossbar_2018, pataki_surface_code_crossbar_2024, kunne_spinbus_2024, escofet2025quantumreversemappingsynthesizing}. Meanwhile in  the Looped Pipeline~\cite{cai_looped_pipelines_2023}  approach both data and ancilla qubits are shuttled around loops to enable the embedding of a 3D stack of error correcting codes in a strictly 2D lattice. As an alternative to all these paradigms where the canonical surface code layout is translated onto a silicon shuttling-enabled lattice, reformatting of logical qubits into 1D objects has also been explored. Refs.~\cite{siegel_two_by_n_2024, micciche2025optimizingcompilationerrorcorrection} examine an implementation of QEC on a 2$\times$N array equipped with shuttling. The more recent Snakes on a Plane architecture~\cite{siegel2025snakesplanemobilelow} showed full-scalability and fault-tolerance when multiple 2$\times$N filaments are interfaced together.

If shuttling is to be used in such an extensive way, it becomes crucial to understand and minimise the noise associated with it. Theoretical studies have demonstrated that one of the main contributions to shuttling noise stems from variations of the spin's $g$-factor while it is shuttled~\cite{Langrock_2023}, resulting in the accumulation of a random phase. A simple idea to mitigate this process is to upgrade every qubit from a mere Loss-DiVincenzo (LD) encoding (using one electron~\cite{lossQuantumComputationQuantum1998}) to a singlet-triplet encoding (ST), now using two electrons whose proper state should lie in the odd-parity subspace~\cite{levy_st_2002}. Interestingly, such qubits are immune against long-correlation length phase-like decoherence processes affecting the pair -- the odd-parity subspace is in this way an approximate \textit{decoherence-free subspace}. For this reason, it is expected that these singlet-triplet, or \textit{dual-spin}, qubits would be more resilient to shuttling noise than LD qubits~\cite{mokeev2024modelingdecoherencefidelityenhancement, siegel2025snakesplanemobilelow, zhang_decoherence_2025}. This may also result in an easier shuttling procedure: even in the absence of randomness from e.g. charge fluctations, LD qubits naturally will pick up a net phase during their shuttle, which must be calibrated away by a full prior characterisation of the $g$-factor landscape.
(indeed this phase shift is strong enough that it has been proposed as a mechanism for single qubit gate rotation~\cite{YonedaCoherentSpin2021}).
ST qubits are naturally first-order immune to this, meaning that this calibration along the whole shuttling track may be unnecessary. Notably, ST qubits can also be fully electrically controlled, which allows for fast and naturally localised gates~\cite{universalFourSingletTriplet}. In comparison, the execution of single-qubit gates in LD qubits requires the application of an oscillatory field, which may have to be global to reduce wiring complexity, and thus  crosstalk must be allowed for~\cite{kane1998,vahapoglu_single-electron_2021, Hansen_2024, Fayyaz_2023}. Also note that, while ST qubits require twice as many electron spins, this does not necessarily translate into a doubled resource overhead. Indeed, spins forming an ST qubit can be close to each other and loaded in a double quantum dot. For an equal number of qubits, LD- and ST-based devices may have comparable sizes, each implementation coming with its own engineering challenges.

\begin{figure*}
    \centering
    \includegraphics[width=\linewidth]{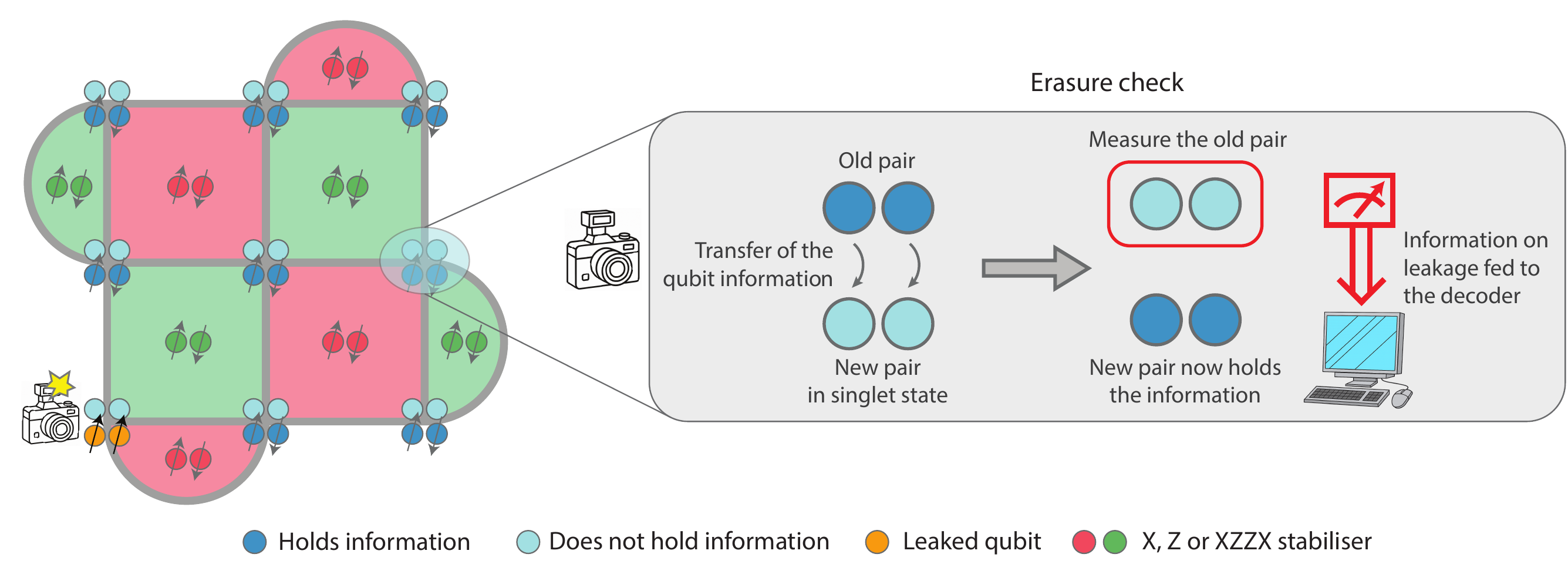}
    \caption{QEC with dual-spin qubits, using the surface code as an example. Left: data and ancilla qubits are encoded in singlet-triplet pairs \textit{i.e.} dual-spins living in the odd-parity subspace. These qubits have a degree of natural protection against phase noise but suffer from leakage errors (orange pair). Stabilisers are not designed to catch such errors thus leakage detection strategies (or erasure checks) must be designed. Right: leakage is checked by unitarily transferring the information from an old to a new data pair before measuring the old one out, which reduces the entropy of the system. This protocol both accurately detects leakage and projects the new pair back to the computational subspace without the need for a measurement feedback.}
    \label{fig:main_figure}
\end{figure*}

These properties therefore make singlet-triplet spin qubits a promising candidate for the implementation of a quantum computer. However, while they do naturally mitigate the impact of certain error mechanisms -- as explained above -- they also suffer from a different kind of noise: leakage. If one spin of a pair flips, the singlet-triplet qubit will leave the odd-parity subspace, therefore leaking from it. This can be catastrophic for an error-corrected quantum computer if these leakage events are not detected and dealt with properly. Indeed standard stabilisers are only designed to catch errors within the computational subspace. If detected however, the leakage events can be leveraged to dramatically increase the performance of the decoder, which is now directly aware of the location of the errors, instead of only inferring them from stabiliser information. There has been extensive research in this direction for various quantum platforms, showing how using \textit{erasure qubits}, \textit{i.e.} qubits whose leakage noise is detectable, can yield significant performance improvements~\cite{gu_fault-tolerant_2025, gu_optimizing_2024}. Erasure conversion, i.e. the advantageous mapping of Pauli-type errors to leakage events, has been investigated in the case of dual-rail photonic qubits~\cite{KLM2001, Varnava2008}, dual-rail superconducting qubits~\cite{kubica_erasure_2023, teoh_dual-rail_2023, wills2025errordetectedcoherencemetrologydualrail}, neutral atoms~\cite{wu_erasure_2022, Sahay_erasure} and ion traps~\cite{kang_quantum_2023}. It is important to understand the benefits of such an approach for semiconductor spin qubits. While the majority of past literature has explored the implementation of QEC with LD qubits~\cite{Takeda_2022, saraiva_dawn_2023, hetenyi_tailoring_2024, jones_logical_2018, gravier2025simulatednonmarkoviannoiseresilience}, a study has  explored switching the ancillas to ST qubits (for ease of measurement, since spin qubits can naturally be initialised in the singlet state and measured in the singlet-triplet basis)~\cite{cai_silicon_2019, gutierrez_comparison_2025}. 

In this paper we take one step further and show how singlet-triplet {\em data} qubits make for excellent erasure qubits. We interchangeably use the term single-triplet (ST) qubit, which is familiar to the spin qubit community, and the term dual-spin qubit which we derive from the common term `dual-rail' used in other modalities. Our methods are summarised in \cref{fig:main_figure}. We begin by analysing a fault-tolerant protocol for erasure detection. Interestingly, we show that it automatically projects the qubits back onto the computational subspace without the need for any feedback loop to the quantum chip: this simplifies classical control and reduces latencies. Then we derive two classes of circuits implementing a stabiliser measurement. The first one relies on exchange gates only and outputs noise that is equal parts Pauli Z, Pauli X and erasure. We evaluate its performance under a standard CSS surface code~\cite{Bravyi_Kitaev_1998, Kitaev_1997, Fowler_sc}. The second class additionally uses electromagnetic driving pulses to implement CNOT gates. While coming at a cost of potentially increased implementation complexity, we show that to first order this approach successfully converts all X noise into erasure, meaning that the leftover undetected Pauli noise is fully biased towards Z. By coupling this to the use of the XZZX surface code~\cite{bonilla_ataides_xzzx_2021}, a variant of the surface code that is tailored to biased noise, we demonstrate even higher error suppression, with a more-than-doubled code threshold, and logical error rates reduced by multiple orders of magnitude.

\section{Physical implementation} \label{sec:physical_implementation}

In this section, we give background on the physical implementation of LD and ST qubits, focusing on the properties of their native operations. A reader that is more interested in the QEC aspects of this research can skip the first two subsections, whose crucial results are summarised in \cref{sec:physical_implementation_summary}.

\subsection{Loss-DiVincenzo qubits}

The LD qubit is the simplest realisation of a spin qubit, where the logical states are encoded in the Zeeman-split spin states of a single confined electron in a semiconductor quantum dot,
\begin{equation}
    \ket{0} = \ket{\uparrow}, \quad \ket{1} = \ket{\downarrow}.
\end{equation}

In an external magnetic field $B_0$, the energy level splitting reads
\begin{equation}
    \omega_q = g \mu_B B_0
\end{equation}
where $g$ is the $g$-factor and $\mu_B$ is the Bohr magneton. Note that we here set $\hbar=1$.

Control of single-spin qubits is typically achieved through electron spin resonance (ESR), by applying an oscillating magnetic field of amplitude $B_1$ and frequency $\omega$ near the Zeeman splitting $\omega_q$. In the frame rotating at the frequency $\omega$, the single-qubit Hamiltonian reads
\begin{equation}
    H_{\text{LD}} = \frac{\omega_q-\omega}{2} \sigma_z + \frac{\Omega}{2}  \sigma_x,
\end{equation}
where $\Omega = g\mu_BB_1$. However, generating local oscillating magnetic fields on-chip is experimentally challenging and limits the scalability of LD qubits. Instead, global fields are generally applied, and put at resonance with the frequency of the targeted qubits. When such an approach is used in a device with many thousands or millions of spin qubits, frequency engineering techniques may therefore required to ensure that spins targetted by the global pulse are on-resonance while those that should not be affected are sufficiently off-resonance. Rapid modification of the qubit's Zeeman energy is possible by sending electrical pulses to Stark shift their $g$-factor \cite{Ferdous2018}.

Two-qubit gates in LD qubits are implemented by turning on the exchange coupling $J$ between two neighbouring dots. Depending on the values of $J$ and of the detuning $\Delta E_Z$ between the dots, and on whether a driving field is applied, this procedure can successfully implement a SWAP, a $\sqrt{\text{SWAP}}$, a CZ or a CROT gate~\cite{burkard_2023}. For instance, CZ gates can be obtained by setting $J=\Delta E_Z/\sqrt{3}$~\cite{Burkard_cz, meunier_cz} and carefully choosing the gate time accordingly. An important thing to note is that exchange gates (without drive) are spin-conserving: this means that even in the presence of errors, a CZ will always preserve $\ket{\uparrow\uparrow}$ and $\ket{\downarrow\downarrow}$ (up to a phase). It can however have a more complex effect on the odd-parity subspace of the pair, spanned by $\ket{\uparrow\downarrow}$ and $\ket{\downarrow\uparrow}$.

Initialisation to the $\ket{0}$ state can be performed by relaxation, or by adiabatically evolving a two-spin ground state. Measurement can be performed via spin-to-charge conversion: for example, when one attempts to force an electron onto an adjacent dot where existing electron(s) are present, the resulting behaviour depends on their mutual spin state due to Pauli exclusion~\cite{burkard_2023}.

\subsection{Singlet-triplet qubits}

An alternative approach is provided by the ST qubit, which encodes information in the joint spin state of two electrons confined in a double quantum dot.
The logical subspace is defined by the spin-singlet and spin-triplet states with total spin projection $S_z = 0$ (odd-parity subspace),
\begin{equation} \label{eq:S_T0_def}
    \ket{S} = \frac{1}{\sqrt{2}}\big(\ket{\uparrow\downarrow} - \ket{\downarrow\uparrow}\big), \quad
    \ket{T_0} = \frac{1}{\sqrt{2}}\big(\ket{\uparrow\downarrow} + \ket{\downarrow\uparrow}\big).
\end{equation}
We will also denote the even-parity states as
\begin{equation}
    \ket{T_-} = \ket{\downarrow\downarrow}, \quad
    \ket{T_+} = \ket{\uparrow\uparrow}.
\end{equation}

In this paper we will define the computational basis states as
\begin{equation} \label{eq:st_basis}
    \ket{0} = \ket{\uparrow\downarrow}, \quad \ket{1} = \ket{\downarrow\uparrow},
\end{equation}
such that $\ket{T_0}=\ket{+}$ and $\ket{S}=\ket{-}$. The equivalent term `dual spin' qubit can be used instead of ST qubit.

In this encoding, single-qubit operations are governed by two tunable energy scales: the exchange coupling $J$, which splits the singlet and triplet states, and a magnetic-field gradient $\Delta B_z$ between the dots, which induces singlet–triplet mixing. The total Hamiltonian of a singlet-triplet qubit thus reads: 
\begin{equation}
    H_{\text{ST}} = \frac{1}{2} \Delta E_Z\, \sigma_z + \frac{J}{2} \sigma_x,    
\end{equation}
with $\Delta E_Z=g \mu_B \Delta B_z$. Both parameters can be controlled electrically — $J$ through the interdot detuning and tunnelling barrier, and $\Delta E_Z$ via \textit{e.g.} Stark shift — enabling all-electrical control of spin rotations without the need for oscillating magnetic fields. Interestingly, Hadamard gates in the computational subspace can simply be implemented by setting $J=\Delta E_Z$. This will be leveraged in the rest of this paper.

\begin{figure}
    \centering
    \includegraphics[width=\linewidth]{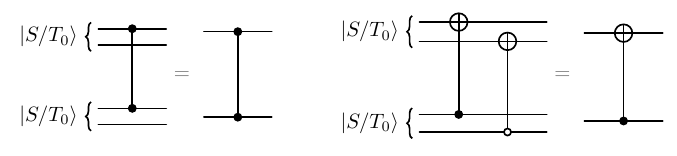}
    \caption{Circuit implementation of the CZ and CNOT gates between two ST qubits.}
    \label{fig:st_gates}
\end{figure}

Two-qubit gates in ST qubits can be applied in multiple ways~\cite{shulman_demonstration_2012, nichol_high-fidelity_2017} but in this paper we will focus on an implementation that is a mere extension of the LD-qubit case. As pictured in \cref{fig:st_gates}, CZ gates can simply be performed by applying a CZ between one spin of each pair. This is because of the encoding we chose in \cref{eq:st_basis}. Similarly to the CZ, a CNOT gate is implemented by applying one CNOT and one anti-controlled CNOT across the pairs. Note that the latter operation can be applied with the same complexity as the regular CNOT gate, by only tweaking the drive frequency (there is no need to apply X gates to the control before and after the gate)~\cite{2q_gate_russ}. An alternative implementation uses the same spin of the controlling pair as the control for both CNOTs. While formally equivalent, this doubles the gate time (as operations cannot be parallelised) and can lead to the generation of harmful errors via propagation through the gates, as will be explored in \cref{sec:stabilisers}.

As explained later under the noise models section (\cref{sec:noise_models}), throughout this paper we assume that gate-induced noise is primarily due to imperfections in the parameters that define the gate (time, voltage, etc). This means gates only give rise to certain types of error — importantly, any operation involving spin-conserving terms will indeed conserve spin, even if noise causes the unitary to differ from the intended one. This applies to the case of exchange-based gates \textit{i.e.} H and CZ. On the one hand, it means that Hadamards cannot cause qubits to leak nor to bring leaked qubits back to the computational subspace. On the other hand, the action of a noisy CZ gate on two ST qubits in the computational subspace cannot leak one qubit without leaking the other, as the total spin of the pair would not be conserved.

For initialisation one might typically prepare a singlet state (for example by pulsing the double dot into a (0,2) charge configuration where both electrons occupy the same dot and relax into the singlet), and then adiabatically convert into the (1,1) charge configuration such that the qubit starts in the known $\ket{S}$ state~\cite{burkard_2023}, i.e. the dual-spin qubit is in the $\ket{-}$ state.

We assume measurement is performed via a spin-to-charge mapping whereby dot potentials are altered to favour two electrons occupying one quantum dot: the timescale for this double occupancy to be reached (relaxation time) depends on the prior spin state in such a fashion that one can determine the most likely prior state. Demonstrations have shown differentiation between $\ket{S}$ and other states (S-versus-T readout), or between the odd-parity space ($\ket{S}$, $\ket{T_0}$) and the even space ($\ket{T_+}$, $\ket{T_-}$) (parity readout)~\cite{seedhouse_pauli_2021}. An advanced measurement device~\cite{lainé2025highfidelitydispersivespinsensing, nurizzo_complete_2023} can be expected to differentiate three cases: $\ket{S}$, $\ket{T_0}$ and ‘other’ (meaning $\ket{T_+}$,$\ket{T_-}$) and we explore the power of such a measurement presently. The probability that the assignment is correct can vary, \textit{i.e.} it may be that the conclusion “the state was $\ket{S}$” is more reliable than a conclusion that “the state was $\ket{T_0}$”, although in the modelling presented here we do not explore such variation. While we will make use of this more elaborate measurement procedure in most of the paper, we also argue in the final section that a weaker version of our protocols can be implemented using only distinct ST and parity readout systems.

\subsection{Summary} \label{sec:physical_implementation_summary}

We here summarise the results of the past subsections that will heavily be used in the rest of the paper.

\paragraph{Singlet-triplet encoding.} The computational subspace is defined by \cref{eq:st_basis}. ST qubits can leak to states $\ket{T_-}=\ket{\downarrow\downarrow}$ and $\ket{T_+}=\ket{\uparrow\uparrow}$.

\paragraph{Pauli error conversion between LD and ST qubits.} We will often work with Pauli errors at the single-spin level (LD qubits). How these errors are ultimately converted into Paulis at the ST level is summarised in \cref{tab:pauli_conversion}.

\paragraph{Operations properties.} The physical properties of the gates we consider imply a structure for the noise (it would be inaccurate to describe it as mere depolarising). Namely, Hadamard and CZ gates are exchange gates hence spin-conserving. This means that even in the presence of errors, the action of H on $\ket{T_\pm}$ is the identity (up to a phase), and it cannot leak an odd state out of the computational subspace:
\begin{align}
    &H\ket{\uparrow\uparrow} \propto \ket{\uparrow\uparrow}, \quad
    H\ket{\downarrow\downarrow} \propto \ket{\downarrow\downarrow} \\
    &H(\text{span}(\ket{\uparrow\downarrow}, \ket{\downarrow\uparrow})) \subset\text{span}(\ket{\uparrow\downarrow}, \ket{\downarrow\uparrow})
\end{align}
As for the CZ gate, it means that even if noisy, it cannot leak one of the qubits it targets without leaking the other. The CNOT gate does not show such properties.

\paragraph{Initialisation.} ST qubits can be initialised in the singlet state $\ket{S} = \ket{-}$ (\cref{eq:S_T0_def}).

\paragraph{Measurement.} ST qubits can be measured in a way that differentiates (ideally) between three outcomes: $\ket{S}$, $\ket{T_0}$ or leaked.

\begin{table}[h]
    \centering
    \begin{tabular}{|c|c|}
        \hline
        Spins & Qubit \\
        \hline
        II, ZZ & I \\
        ZI, IZ & Z \\
        XX, YY & X \\
        XY, YX & Y \\
        \hline
        \makecell{IX, XI, IY, YI,\\ XZ, ZX, YZ, ZY} & \makecell{L} \\ 
        \hline
    \end{tabular}
    \caption{Conversion table between Pauli errors on the individual spins of an ST pair, and the resulting Pauli operation on the ST qubit. L indicates a leakage outside the computational subspace.}
    \label{tab:pauli_conversion}
\end{table}

\section{Fault-tolerant protocols for singlet-triplet spin qubits}

In this section, we present all the strategies we designed for implementing a fault-tolerant quantum computer based on singlet-triplet spin qubits. We focus on the surface code in memory mode only, although the implementation of logical gates can easily be deduced from the native operations presented in the previous section. 

The underlying idea is that the singlet-triplet pairs can be viewed as a base code -- an error-detect bit-flip code -- that we concatenate to the surface code. Single-spin bit flips within an ST pair correspond to a leakage out of the computational subspace, and can be detected via the protocol described in \cref{sec:leakage_detection}. Efficient stabilisers circuits are implemented in \cref{sec:stabilisers}. The knowledge of aforementioned flips can then be fed to a tailored decoder as explained in \cref{sec:decoder}.

\subsection{Leakage detection} \label{sec:leakage_detection}

\begin{figure}
    \centering
    \includegraphics[width=\linewidth]{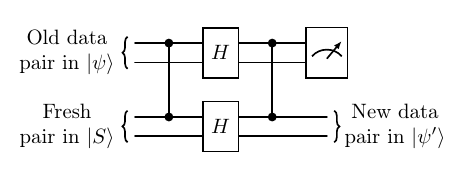}
    \caption{Leakage detection circuit and projection back onto the computational subspace. When the initial state $\ket{\psi}$ is in the computational subspace, its measurement yields a singlet and the output state $\ket{\psi'}$ is equal to $X\ket{\psi}$. The X flip is not problematic as it is deterministic and can be stored in memory. When the input state is leaked however, the measurement outcome becomes $\ket{T_\pm}$ and the output $\ket{\psi'}$ is in an (unknown) state back in the computational subspace.}
    \label{fig:leakage_detection}
\end{figure}

We here present our leakage detection protocol, which additionally projects the qubits back onto the computational subspace without the need for any feedback loop or additional operation. As illustrated in \cref{fig:main_figure}, the idea is to map the state of each data qubit onto a fresh singlet pair and measure out the old pair. While data qubits should in general never be measured in a conventional QEC setup, the operation is here permitted as the state of the pair is not collapsed if it is in the computational subspace. One can view this procedure as measuring the parity checks of the aforementioned base code -- the error-detect bit-flip code. We give a circuit implementation of this protocol in \cref{fig:leakage_detection}. The old data pair (to be assessed) is in a general state $\ket{\psi}$ of the two-spin Hilbert space, and the fresh pair is prepared in the singlet state. The first steps consist of unitarily mapping the old state onto the new pair via four exchange gates (setting $J=\Delta E_Z$ for H and $J=\Delta E_Z/\sqrt{3}$ for CZ). Then the old pair is measured with the measurement protocol we described in the last paragraph of \cref{sec:physical_implementation_summary}. In the absence of errors, an input state $\ket{\psi}$ in the computational subspace is perfectly transferred onto the new pair (up to a deterministic Pauli X that can be regarded as a known frame update), and the measurement outcome is a singlet. If the input was leaked however, the measurement flags it by returning $\ket{T_\pm}$. As mentioned above, the output $\ket{\psi'}$ is directly in an (unknown) state in the computational subspace, without any need for additional gates to project it back. This arises from our choice of gates, which necessarily preserve the parity of the singlet state. We show detailed calculations proving this in Appendix \ref{app:leakage_detection}.

This protocol thus exactly performs what we require: it preserves states in the computational subspace, but projects leakage back without the need for feedback to the quantum chip. This is achieved by measuring out the data qubit, thereby reducing the entropy of the system. These scenarios are distinguished by different measurement outcomes, which enables the use of an erasure-aware decoder.

\subsection{Stabiliser circuits} \label{sec:stabilisers}

\begin{figure}
    \centering
    \includegraphics[width=\linewidth]{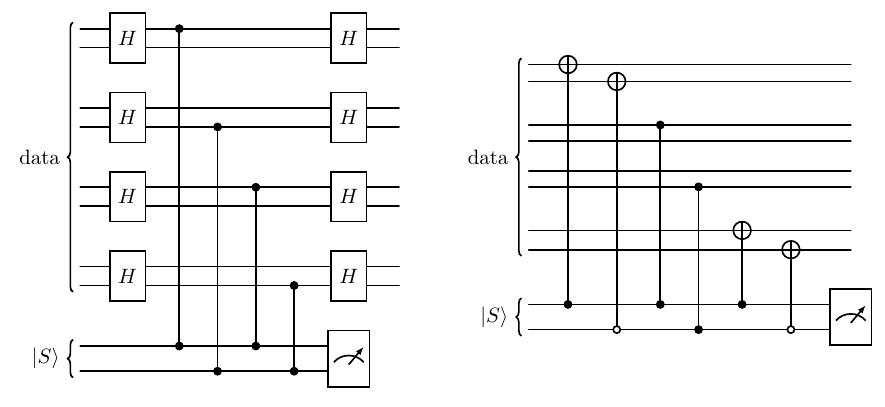}
    \caption{Stabiliser circuits for the ST qubits. Each pair of lines represents one ST qubit. The measurement is assumed to distinguish between $\ket{S}$, $\ket{T_0}$ and $\ket{T_\pm}$ as explained in \cref{sec:physical_implementation}. Left: X stabiliser using exchange gates only. The Z stabiliser is obtained by removing the H gates. Right: XZZX stabiliser, using exchange and driven gates. This stabiliser is bias-preserving.
    }
    \label{fig:stabilisers}
\end{figure}

By performing the leakage detection circuit between every stabiliser round, one makes sure that the state is almost always in the computational subspace (as it is naturally projected back). In this section, we design stabiliser circuits, required to catch errors within the computational subspace. Given the variety of native gates that semiconductor spin qubits possess (see \cref{sec:physical_implementation}), there is some liberty for the stabilisers implementation. We will introduce two families of such circuits: one that only uses exchange gates and one that uses both exchange and driven gates. They are represented in \cref{fig:stabilisers}. In the left panel, we show the implementation of an X stabiliser (or Z if the H gates are removed), required for the standard CSS surface code. This only relies on exchange gates (setting $J=\Delta E_Z$ for H and $J=\Delta E_Z/\sqrt{3}$ for CZ). This implementation is therefore attractively simple, as all operations are local and electrically-controllable.

Looking at the circuit more closely, one can note that using ST rather than LD qubits effectively converts some Pauli X noise into erasure noise, which is detectable in virtue of the protocol presented in \ref{sec:leakage_detection}. Indeed, if one of the entangling gates was followed by an X error, it would only affect one spin of the pairs it is acting on. However, a Pauli X for an ST qubit requires an X and/or a Y error on both spins of the same pair (as per \cref{tab:pauli_conversion}). A single X is instead a leakage event. Subsequent entangling gates then offer no opportunity for the error to propagate to the other spin of the pair and the final Hadamards preserve leakage (as noted in \cref{sec:physical_implementation}). This proves the partial erasure conversion properties of this circuit.

While this is already valuable, one can go even further and get rid of all first-order Pauli X noise. This was not the case before as Z errors can occur after a CZ, which would then be converted into Pauli X by the final Hadamard gates -- they are not bias-preserving. Instead, one can make use of CNOT gates in place of all groups of H-CZ-H. We represent this in the right panel of \cref{fig:stabilisers} for an XZZX stabiliser (which by definition measures the X parity of qubits 1 and 4, plus the Z parity of qubits 2 and 3)~\cite{bonilla_ataides_xzzx_2021}. We use \cref{fig:st_gates} for the circuit design. With the same reasoning as before, an individual entangling gate and subsequent error propagation cannot generate a Pauli X error as this requires a flip of both spins of an ST pair. This cannot happen as we alternate which spin of the ancilla pair controls the CNOT gates. Therefore, to first order this stabiliser circuit only gives rise to Pauli Z and leakage noise. This is particularly powerful as the XZZX surface code has a threshold that is almost twice as high as that of the standard CSS surface code in the infinite-bias limit~\cite{bonilla_ataides_xzzx_2021}. Note however that we do have to implement the leakage detection circuit shown in \cref{fig:leakage_detection}, which can trigger Pauli X errors because of the use of Hadamard gates. It however features properties that we can exploit to retrieve this higher bias, as explained in the next section.

Finally, note that, thanks to the use of dual-spin qubits and to the structure of the noise, hook errors are not a problem here. This term refers to errors arising after a given gate, then propagating to more than one data qubit via subsequent entangling gates. This can lead to a reduced effective distance of the code if the ordering of the gates is not chosen wisely. We achieve this by purposely alternating which spin of the ancilla pair is used as control in \cref{fig:stabilisers}, and provide a detailed explanation in Appendix \ref{app:hook_errors}.

Likewise, a common problem with erasure qubits is the propagation of leakage from one data to another via an ancilla qubit. We avoid this here by always using the ancilla as the control in the stabiliser circuit's entangling gates. This way, leakage -- characterised by an X or Y flip of a single spin of a pair -- cannot propagate to the ancilla (only Z errors can).

\subsection{Decoder} \label{sec:decoder}

One of the main characteristics of the ST encoding compared to the standard LD approach is that qubits can leak out of the computational subspace. In \cref{sec:leakage_detection} we explained how to detect such leakage. In this section we explore how this information can be fed to the decoder. Although decoders such as Union Find can be superior choices for erasure noise~\cite{Delfosse_uf}, here we opt for Minimum-Weight Perfect Matching (MWPM) as it is attractive for its simplicity in this initial study. The principle of this decoder is to build a matching graph whose vertices are the spacetime syndrome generated by the errors. The weight between two vertices is related to the probability that an error mechanism generated them. MWPM then finds the lowest-weight correction explaining the syndrome.

\subsubsection{CSS surface code}

Let us first focus on the simpler case of the standard CSS surface code. In this work, we set the edge weight between neighbouring stabilisers in spacetime to 1 for the CSS surface code. This is sub-optimal under the circuit-level noise models we will consider thereafter but simpler to implement. The complexity and advantage of our decoder arises from the treatment of leakage.

Let us first focus on the case of the data qubits. We adopt a strategy similar to~\cite{Xu_sc_concatenated} whereby the edges weights in the matching graph are adjusted according to detected erasure. More precisely, we decide to set the weight of all edges passing through a leaked data qubit to 0 (there is a $50\%$ chance that such a qubit is faulty after being projected back to the computational subspace). In the absence of errors in the leakage detection circuit, the leaked qubit's locations are perfectly known: measuring a $\ket{T_\pm}$ on the old pair means that it was leaked, thus that the new pair is in a random state within the computational subspace. However, first-order error mechanisms such as measurement errors can wrongly deem a given data qubit as faulty. To circumvent this, we can exploit the spin-conservation properties of the exchange gates. Effectively, there are only two first-order scenarios whereby a given data qubit can leak out of the proper subspace. Note that a Hadamard gate cannot cause a data qubit to leak. However, a CZ gate can cause both of the two pairs it is acting on to leak at the same time (not one or the other, see \cref{sec:physical_implementation_summary}). If the faulty CZ is in a stabiliser circuit, this means that both a data and an ancilla qubits would be leak from the proper subspace. To first-order, leakage will then then be detected by both the ancilla measurement and the subsequent leakage detection circuit. 

\begin{figure}
    \centering
    \includegraphics[width=\linewidth]{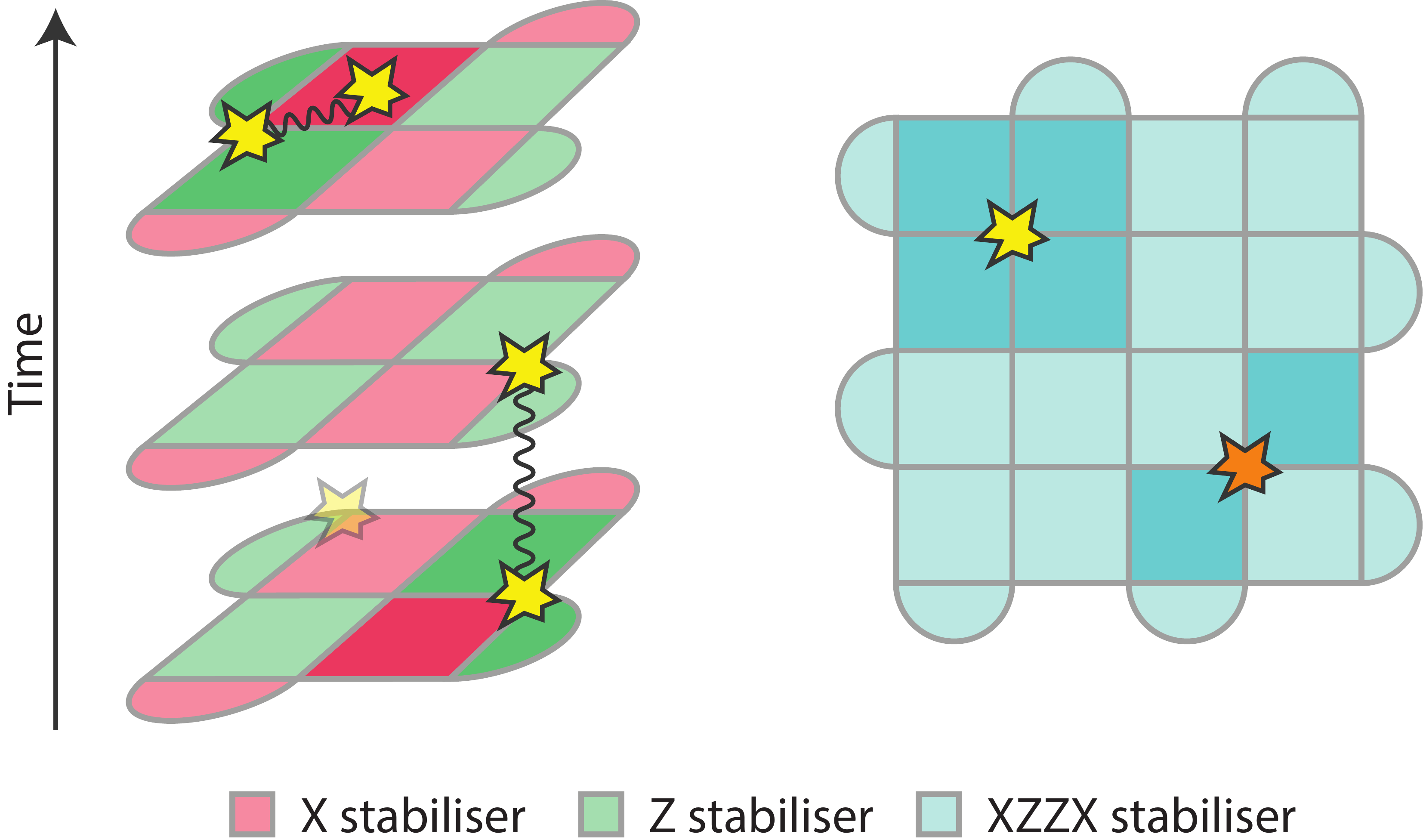}
    \caption{Decoding is performed via MWPM, whose weights are adjusted according to the additional information extracted from the leakage detection circuit. Left: standard CSS surface code. Leakage detections correspond to the measurement of a $\ket{T_\pm}$ state (yellow stars). The exclusive use of exchange gates allows one to confirm leakage by pairing such detection events in space or time (see main text). Isolated detections (faded) are most likely the result of measurement errors. In case of leakage, both neighbouring X and Z stabilisers (in bright colours) are connected with weight 0. Right: XZZX surface code. We consider both $\ket{T_\pm}$ and $\ket{T_0}$ as detection events (resp. yellow and orange stars). The latter is used to detect Pauli X errors only (therefore only the weight between the SW and NE stabilisers is set to 0). Detections here cannot be paired due to the use of CNOTs, which are not spin-conserving, thus there is a higher risk of false leakage detection.}
    \label{fig:decoder}
\end{figure}

If however the faulty CZ is in the leakage detection circuit itself, then both the old and new data pairs would be caused to leak. To first-order, leakage would thus be detected by the current and next leakage detection circuits. Therefore, one can see that first-order leakage detections always come in pairs. By verifying if a leakage detection event has a neighbour in spacetime, one can significantly increase the erasure detection capability. We represent this in the left panel of \cref{fig:decoder}.

As for the ancilla qubits, we follow a simpler procedure. In a standard QEC setup, the syndrome is calculated by taking the difference between the stabiliser measurements at two consecutive rounds. This enables the differentiation between measurement errors and data-qubit-error detections. This cannot directly be generalised in our case since the ancilla qubit's measurements can yield three distinct outcomes. To go back to the previous case, our strategy consists in \textit{pasting over} the stabiliser's values from the previous and next rounds. Using $s_t$ to denote the stabiliser measurement at time $t$, then if $s_t=T_\pm$ we act thus:
\begin{itemize}
    \item If $s_{t-1}=s_{t+1}$: set $s_t=s_{t-1}=s_{t+1}$
    \item Else: randomly set $s_t=S$ or $s_t=T_0$ with $50\%$ chance.
\end{itemize}
This simple intuitive rule erases the ancillas' leaked measurements (after their information was used for the data qubits erasure detection, as explained above). After this step the regular difference syndrome can be computed.

\subsubsection{XZZX surface code}

For the XZZX surface code, the approach is slightly different. We first set three distinct weights $w_X$, $w_Z$ and $w_t$, accounting for the bias in the observed noise~\cite{bonilla_ataides_xzzx_2021}:
\begin{equation} \label{eq:xzzx_weights}
    w_Z = w_t = \log\left(\frac{1-p}{p}\right) ~~ \text{and} ~~
    w_X = \log\left(\frac{1-p^2/10}{p^2/10}\right)
\end{equation}
where $p$ is the common gate infidelity (see \cref{sec:noise_models}). The choice of $p^2/10$ is estimated from simulations showing that when the Z noise is $\Theta(p)$, the X noise is $\Theta(p^2/10)$ if only accounting for errors in the stabiliser circuit. The second order is backed by the arguments of the previous section, proving that the circuit in the right panel of \cref{fig:stabilisers} converts all first-order Pauli X errors into leakage. The division by 10 approximately accounts for this.

However, as noted earlier, the leakage detection circuit of \cref{fig:leakage_detection} does \textit{not} show such a high bias. Yet, by analysing the possible first-order Pauli errors that can affect the output state $\ket{\psi'}$ after propagation through the gates, one can show that there are only two scenarios giving rise to a Pauli X in the computational subspace. They correspond to a Z or a ZZ error after the first CZ, and propagate as follows:
\begin{align*}
    \text{IIZI} &\xrightarrow{HH} \text{IIXX} \xrightarrow{CZ} \text{IZXX} \\
    \text{IZZI} &\xrightarrow{HH} \text{XXXX} \xrightarrow{CZ} \text{XYYX}
\end{align*}
where each character corresponds to the error on each given spin. Here we have used the fact that a Hadamard converts a Pauli Z in the computational subspace \textit{i.e.} ZI or IZ on the spins, into a Pauli X \textit{i.e.} XX or YY (see \cref{tab:pauli_conversion}). Applied to the perfect output $\ket{S}\ket{\psi}$, one respectively obtains $\ket{T_0}XX\ket{\psi}$ and $\ket{T_0}YX\ket{\psi}$. This means that X (and Y) errors in the computational subspace can only occur when a triplet $\ket{T_0}$ is measured. While this measurement is not restricted to the occurrence of such a bit flip (it can \textit{e.g.} happen with a simple measurement error), it however indicates that to first-order \textit{all} X errors are at locations where a $\ket{T_0}$ was measured. For this reason, we set the weight of X edges where an old data pair was measured in $\ket{T_0}$ to 0. This is represented with the orange star in the right panel of \cref{fig:decoder}.

As for the treatment of leakage, we adopt a simpler strategy than earlier as CNOT gates do not possess the same spin-conservation properties as the CZs. Therefore one cannot detect all first-order erasures by pairing leakage detection events. We thus resort to declaring that leakage happened wherever the measurement of an old data pair shows $\ket{T_\pm}$, although this could clearly also be the result of a measurement error. We set corresponding edges weights to 0 in such a case (both X and Z edges however, see e..g the yellow star in the right panel of \cref{fig:decoder}). As for ancillas which are found to have leaked when measured, we use the same pasting over strategy as for the CSS surface code.

\section{Surface code simulations}

The previous section introduced all the protocols that we wish to study. Here we apply them to the simulation of a square CSS or XZZX surface code in memory mode and confirm the high performance of our methods.

\subsection{Noise models} \label{sec:noise_models}

Let us describe the noise models we adopt, first focusing on gate noise. While many QEC papers consider a generic depolarising channel, we here design a noise model that is more faithful to the operations that are implemented in the device. For instance, in the leakage detection pairing algorithm presented in \cref{sec:decoder}, we crucially used the fact that imperfect H gates cannot generate leakage, and that imperfect CZs can only cause {\em both} qubits they are acting on to leak at the same time. Therefore, modelling their noise with a generic depolarising channel would not be correct. Instead, we model the lower-level gate noise with random variations of the physical parameters: the exchange coupling $J$, the Zeeman energy difference between the spins $\Delta E_Z$ and the gate duration $T$. These uncontrolled fluctuations can arise from charge noise~\cite{kepa_2023, elsayed_low_2024, yoneda_noise_correlation_2023, jnane_ab_initio_2024, shehata_2023, yonedaQuantumdotSpinQubit2018, culcer_2009, rojas_spatial_correlations_2023} or imperfect classical control. For a given imperfect set of such parameters, we obtain a faulty unitary evolution $U$. We can then write the average gate fidelity over any initial state with respect to the perfect unitary $U_\text{target}$~\cite{pedersen_fidelity_2007}:
\begin{equation} \label{eq:fid_formula}
    \mathcal{F}(J,\Delta E_Z,T) = \frac{d+|\text{Tr}(U^\dagger_\text{target}U)|^2}{d(d+1)}
\end{equation}
Then we extract the average fidelity over the distribution $p(J,\Delta E_Z,T)$ of physical parameters:
\begin{equation}
    \overline{\mathcal{F}} = \iiint \mathcal{F}(J,\Delta E_Z,T)p(J,\Delta E_Z,T)~\mathrm{d}J \mathrm{d}\Delta E_Z \mathrm{d}T.
\end{equation}
We assume that the physical parameters are independent and follow a Normal distribution:
\begin{equation}
    p(J,\Delta E_Z,T) = p_J(J)p_{\Delta E_Z}(\Delta E_Z)p_T(T)
\end{equation}
where each single-parameter distribution is Gaussian \textit{e.g.}
\begin{equation}
    p_T(T) = \frac{1}{\sqrt{2\pi}\sigma_T} \mathrm{e}^{-(T-T_0)^2/2\sigma_T^2}
\end{equation}
where $T_0$ is the perfect gate time and $\sigma_T$ the standard deviation.
In the QEC simulations, we will be setting the average fidelity $\overline{\mathcal{F}}$ to \textit{e.g.} $99.9\%$. To reproduce this value, we need to determine the values of the standard deviations achieving $\overline{\mathcal{F}}$. Once this has been evaluated, we sample a set of parameters $(J,\Delta E_Z,T)$ for each gate location and fix it for the rest of the simulations. This accounts for the fact that the gate's properties are defined by their surrounding electromagnetic environment and assumes that it is static over time. Ultimately, this means that each gate location has a fixed fidelity $\mathcal{F}$, but that only the average $\overline{\mathcal{F}}$ over all gates was initially given.

In this paper we will only model the influence of the timing errors as it leads to simple analytical expressions between $\overline{\mathcal{F}}$ and $\sigma_T$:
\begin{align}
    \overline{\mathcal{F}}_{\text{1QLD}} &= 1-\frac{(\Omega_\text{1QLD}\sigma_\text{1QLD})^2}{6} \label{eq:fid_analytic_expressions1} \\
    \overline{\mathcal{F}}_{\text{H,CZ}} &= 1-\frac{(\Omega_\text{H,CZ}\sigma_\text{H,CZ})^2}{10} \label{eq:fid_analytic_expressions2} \\
    \overline{\mathcal{F}}_{\text{CNOT}} &= 1-\frac{(\Omega_\text{CNOT}\sigma_\text{CNOT})^2}{2} \label{eq:fid_analytic_expressions3}
\end{align}
where we differentiate the strength of the timing errors for each gate family: 1QLD for single-qubit gate on LD qubits (implemented with a global driving field); H,CZ for exchange gates between two spins; CNOT for a CNOT gate between two spins. The different $\Omega$'s are the gate frequencies. Calculations are given in Appendix \ref{app:noise_models}.

Although we only present results corresponding to gate noise originating from timing imperfection, we did run additional simulations confirming that the influence of the other parameters would have a similar impact on the numerical results. In particular, they all lead to the same noise structure \textit{i.e.} the same number of non-zero terms in the matrix of the imperfect unitary $U$. The sum of their contribution then leads to the fidelity $\overline{\mathcal{F}}$ which we set. Details can be found in Appendix \ref{app:noise_models}.

On top of gate noise, we will also consider the influence of shuttling and measurement noise. Here we will model these two processes with simple unital channels. Namely, we suppose that both the data and ancilla qubits are shuttled between each two-qubit gate operation (as it is the case for \textit{e.g.} the spiderweb or the looped pipeline architectures~\cite{boterSpiderwebArraySparse2022, cai_looped_pipelines_2023}). This is supposedly advantageous as shuttling may lead to an increased effective dephasing time $T_2^*$, meaning that leaving a qubit idle would be detrimental compared to shuttling it~\cite{Langrock_2023}. Shuttling both data and ancilla would also halve the shuttling distance, thereby reducing shuttling times. We denote the shuttling error per two-qubit gate as $p_\text{sh,LD}$ or $p_\text{sh,ST}$, and model it with a dephasing channel. As explained in the introduction, it is expected that ST qubits would be less prone to shuttling-induced dephasing errors than LD qubits, meaning that $p_\text{sh,ST} \leq p_\text{sh,LD}$.

As for measurement errors, in the case of LD qubits we assume a classical flip of the measurement outcome with a probability $p_\text{meas,LD}$. For ST qubits, as the measurement process can yield three distinct outcomes, we suppose that it leads to any of the two erroneous ones with probability $p_\text{meas,ST}/2$.

\subsection{Simulations details}

The noise models we adopted lead to coherent gate errors. In order to simulate sufficiently large code sizes, one needs to convert these into probabilistic Pauli errors so as to enable the use of an efficient classical simulator. For this purpose, the twirling approximation is widely used in the literature; this can be justified by noting that noise is decohered by the stabiliser measurements for large codes~\cite{Bravyi_2018}. Assuming this, we can replace all coherent noise sources with unital noise channels \textit{e.g.} a coherent phase gate $R_z(\theta)$ would be replaced with a dephasing channel of probability $\sin^2(\theta/2)$.

This assumption enables the efficient simulation of standard QEC experiments, which only make use of Clifford gates. However, this is not the case in our setup. First, the noisy gates we consider are coherent rotations beyond simple Paulis, Hadamards or CNOTs, and we only wish to apply the twirling approximation \textit{after} the stabiliser and leakage detection measurements. In other words, we want to study how coherent errors can propagate within a stabiliser cycle, and only make use of the twirling approximation before decoding the information. This describes noise propagation more accurately. Second, even in the absence of noise, the Hadamard gate that we use is only a Hadamard in the common sense when applied to qubits in the computational subspace. For leaked qubits it is equivalent to the identity gate (\cref{sec:physical_implementation_summary}): this is therefore not a Clifford gate.

For these reasons, we do not use conventional QEC simulators such as $\texttt{stim}$. We instead designed an in-house QEC pipeline which describes our system more accurately. More details can be found in Appendix \ref{app:simulations_details}.

The general workflow is the following: for a given choice of code and encoding (LD or ST, CSS or XZZX), as well as fidelities and code distance $d$, we first sample the local physical parameters $(J,\Delta E_Z,T)$ at each gate location. We then run between 10,000 and 10,000,000 Monte-Carlo samples (depending on the target logical error rate), where errors are randomly applied to data and ancilla qubits for $d$ rounds of stabiliser and leakage measurements. For every run, the information from both these measurements is fed to the decoder to compute a correction and determine if an X or Z logical error occurred. We then deduce a logical error rate $P_L$ \textit{i.e.} the probability that either an X or Z logical failure occurs. Note that we do not decouple the decoding of the Paulis X and Z as the XZZX surface code is non-CSS.

\subsection{Results} \label{sec:results}

\subsubsection{Threshold estimation}

\begin{figure}
    \centering
    \includegraphics[width=\linewidth]{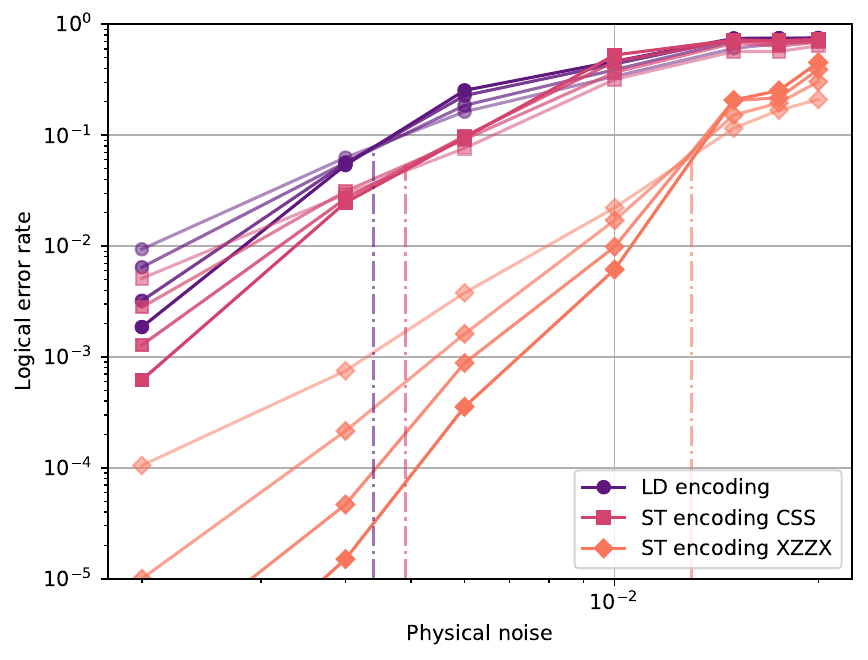}
    \caption{Logical error rate $P_L$ against physical noise $p$ for the three different approaches considered in this paper. Shuttling noise is here set to 0 while all other noise mechanisms have a common infidelity $p$. Each colour corresponds to a different approach, and each line to a given code distance $d\in\{7,9,11,13\}$. The crossing points between same-colour lines is the threshold, highlighted with the dashed lines. Its values for LD, CSS+ST and XZZX+ST respectively are $0.45\%$, $0.49\%$ and $1.3\%$.}
    \label{fig:threshold}
\end{figure}

We first aim to extract the threshold associated with the two approaches presented in this paper. We also simulate the more standard LD encoding and use it as a reference. Results are presented in \cref{fig:threshold}, where the infidelity of all noise mechanisms except shuttling is set to a common value $p$. We do not consider shuttling noise in this plot as its impact on LD and ST is different. One can see that both the CSS and XZZX approaches achieve lower logical error rates and higher thresholds (resp. $0.49\%$ and $1.3\%$) than the LD one ($0.45\%$). This is not something that could have have been predicted from the start, as, while our erasure conversion protocols are strong, errors may be introduced at \textit{every} round and on \textit{every} data qubit by the leakage detection circuits. XZZX stabilisers also use more gates (hence are more prone to errors) as every CNOT between ST qubits is implemented with two physical CNOTs (\cref{fig:st_gates}). The performance of the latter scheme is notably higher than both others, which can be explained with the advantageous error reduction capabilities of the XZZX surface code in the presence of highly-biased noise. This bias was achieved by both the choice of the stabiliser circuit and the Pauli X detection scheme presented in \cref{sec:decoder}. One can additionally note the higher gradient of the lines for this scheme, reflecting the greater effective distance of the XZZX surface code under high bias.

While the CSS variant does not achieve such a high performance, we envision a simple improvement where leakage detections are not performed at every round. Instead, one could decide to only implement them on data qubits that neighbour a leaked ancilla -- further exploiting the pairing property of \cref{fig:decoder}. This would reduce number of physical errors introduced while only moderately affecting the leakage detection accuracy.

As for the XZZX scheme, it seems that the performance is partially limited by the false detection of leakage events (since they cannot be pair-matched as in the CSS case, due to the use of CNOTs). One way to increase the leakage detection accuracy is to account for the natural asymmetry in the measurement error. For example, if the perfect measurement outcome was $\ket{T_0}$, we supposed that there was an equal chance of measuring $\ket{S}$ or $\ket{T_\pm}$. However, as these two outcomes have extremely different relaxation times, it is highly unlikely that they would be confused for one another~\cite{lainé2025highfidelitydispersivespinsensing}. This can be capitalised on at the decoding stage to better distinguish between no error, leakage and Pauli X detection.

Also note that the relatively low threshold for the standard LD encoding (below the canonical circuit-level surface code threshold of $0.7\%$) was to be expected. In our simulations we made use of more detailed noise models including coherent errors, beyond the uncorrelated depolarising that is typically assumed. Additionally, the performance of all approaches is negatively impacted by the simple decoder heuristic we used, where edge weights in the matching graph are not set from individual error mechanisms (set to 1 for CSS or by \cref{eq:xzzx_weights} for XZZX). With comprehensive tuning of the decoder, or the adoption of alternative decoder paradigms, we should expect a collective improvement of all the schemes.

\subsubsection{Impact of shuttling errors}

\begin{figure}
    \centering
    \includegraphics[width=\linewidth]{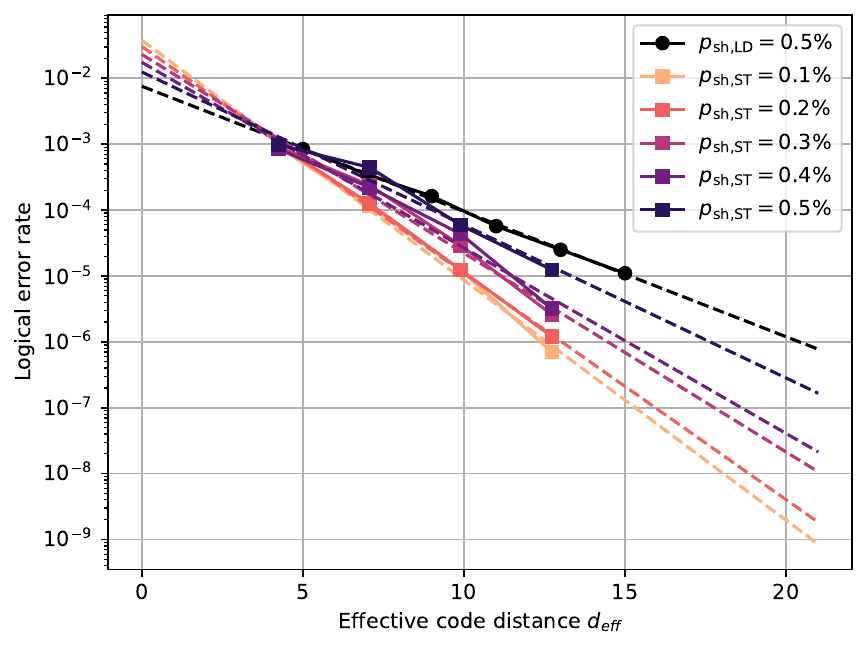}
    \caption{Logical error rate $P_L$ against the effective code distance $d_{\text{eff}}$ for various levels of shuttling noise. We compare the standard LD encoding (black, $d_{\text{eff}}=d$) with the ST+XZZX protocol ($d_{\text{eff}}=d/\sqrt{2}$ as this encoding requires twice as many spins). In both cases, we set the infidelities of all noise mechanisms but shuttling to $p=0.1\%$. We choose $p_\text{sh,LD}=0.5\%$ and vary the level of shuttling noise on ST qubits $p_\text{sh,ST}$ between $0.1\%$ and $0.5\%$. Dashed lines are linear regressions of computed data points.}
    \label{fig:shuttling}
\end{figure}

Given the demonstrated higher performance of the ST+XZZX scheme, we now solely focus on comparing this protocol to the LD one. In \cref{fig:shuttling}, we turn shuttling errors back on and plot the logical error rate $P_L$ against an effective code distance $d_\text{eff}$ for various values of the shuttling noise. This effective distance is simply equal to $d$ (the normal code distance) for the LD encoding and $d/\sqrt{2}$ for the ST encoding (since this encoding uses twice as many spins). This way, we are comparing both schemes at an equal number of spins rather than qubits. We set all other noise sources to $p=0.1\%$. Only the value $p_\text{sh,LD}=0.5\%$ is plotted for the LD encoding, which we compare to values of $p_\text{sh,ST}$ ranging between $0.1\%$ and $0.5\%$ (thus keeping $p_\text{sh,ST} \leq p_\text{sh,LD}$). These parameters quantify the severity of shuttling errors occurring between consecutive two-qubit gates, on both the data and ancilla qubits. Using demonstrated shuttling fidelities from Ref.~\cite{desmet2024highfidelitysinglespinshuttlingsilicon}, $p_\text{sh,LD}=0.5\%$ translates into having to shuttle both data and ancilla across 10$\mu$m, which should more than suffice for any of the spin-shuttling-based fault-tolerant architectures present in the literature~\cite{boterSpiderwebArraySparse2022, li_crossbar_2018, pataki_surface_code_crossbar_2024, kunne_spinbus_2024, escofet2025quantumreversemappingsynthesizing, cai_looped_pipelines_2023, siegel_two_by_n_2024, micciche2025optimizingcompilationerrorcorrection, siegel2025snakesplanemobilelow}.

One can again observe the higher performance of the ST encoding combined with an XZZX surface code, even in the unlikely regime $p_\text{sh,ST} = p_\text{sh,LD}$. When $p_\text{sh,ST}$ is a fraction of $p_\text{sh,LD}$, we show orders of magnitude reduction in the logical error rate for ST compared to LD qubits.

Note that we here compared LD and ST schemes at equal numbers of electrons, which is a most likely a worst-case scenario. Indeed, as noted in the introduction, using two spins per ST qubits cannot directly be thought of as a doubled resource overhead, as these have to be loaded in a double quantum dot thus very close to each other. One would expect a non-trivial adjustment in the device size, the number of measurement components, and the extent of control electronics. Indeed, one can even argue that the XZZX approach would help reduce the overhead as the code can be shrunk in the X direction, given the high error bias~\cite{bonilla_ataides_xzzx_2021}.

\section{Discussion}

In this paper, we propose for the first time a fault-tolerant framework for quantum error correction based entirely on singlet–triplet (dual-spin) qubits, which we identify as a natural realisation of erasure qubits in semiconductor platforms. Building upon the growing body of work leveraging electron shuttling for scalable architectures, we have argued that replacing single-spin (Loss–DiVincenzo) data qubits with dual-spin encodings can substantially improve resilience to shuttling-induced noise, thanks to their intrinsic protection within their decoherence-free subspace. While this upgrade introduces a new error mechanism, namely leakage out of the computational subspace, we design a hardware-efficient erasure-detection circuit. Interestingly, we show that it naturally forces the qubit back onto the computational subspace without the need for a measurement feedback, reducing classical control overheads and timing constraints.

By analysing two families of stabiliser measurement circuits, we demonstrate how the error channels of the resulting operations can be engineered: $(i)$ an exchange-only protocol where erasures occur in pairs and so can be identified with remarkable accuracy, and $(ii)$ a driven-gate protocol converting most X-type noise into erasure, thereby creating a strongly biased noise model (at the cost of a slightly lower leakage detection performance). This only requires gates that are conventionally used for LD qubits. When combined with the XZZX surface code and bespoke leakage-aware decoders, the latter noise tailoring leads to a twofold increase of the threshold and several orders of magnitude improvements in logical error rates.

These results therefore highlight that the singlet-triplet encoding is not a mere noise-resilient upgrade to existing semiconductor architectures. It is instead a pathway toward erasure-based quantum error correction. With continued advances in singlet–triplet control, coupling, and shuttling, this approach brings fully fault-tolerant, semiconductor-based quantum computation significantly closer to realisation.

As future work, we first envisage the use of a higher-performing decoder. We chose a relatively straightforward implementation that is blind to the specific probabilities of each individual error mechanism. This was because our stabiliser circuits include coherent errors and non-Clifford gates, preventing us from using well-established tools such as \texttt{stim}. We expect that upgrading our decoder would collectively improve the performance of all the schemes presented here. Leakage could also be treated less heuristically, by setting the weight of leaked edges to some other value than zero. Further, integrating a Union-Find decoder may lead to even higher improvements.

Additionally, it would be valuable to quantify what precisely limits the performance of each scheme. For instance, one could find a trade-off between measuring the leakage detection circuits at each round, which introduces more physical errors, or more scarcely, which degrades the leakage detection. As noted in \cref{sec:results}, using the gate's symmetries may help identify when these erasure checks should be applied, rather than systematically performing them.

Finally, we remark that we have made extensive use of a measurement procedure that is able to distinguish between $\ket{S}$, $\ket{T_0}$ and $\ket{T_\pm}$. While advantageous, we stress  that this is not strictly required for the implementation of our protocols. Indeed, one could replace ancilla measurements with a more conventional ST readout (which distinguishes $\ket{S}$ from $\ket{T_0}$), and data measurements (in the leakage detection gadget) with a parity readout (which distinguishes $\ket{S/T_0}$ from $\ket{T_\pm}$). This would still satisfy the main requirements of stabilisers and leakage measurements \textit{i.e.} catching errors respectively within and outside the computational subspace. What would be lost is the detection events pairing (CSS scheme) and the Pauli X detections (XZZX scheme). An interesting line of research would be to evaluate the degradation in performance in this case.

\section{Acknowledgements}
The authors are very grateful to Hamza Jnane for several very useful conversations. The numerical modelling involved in this study made use of the Quantum Exact Simulation Toolkit (QuEST) \cite{jones_quest_2019}, and the recent development pyQuEST \cite{pyquest} which permits the user to use Python as the interface front end. We are grateful to those who have contributed to these valuable tools. The authors acknowledge support from EPSRC’s Robust and Reliable Quantum Computing (RoaRQ) project (EP/W032635/1).

\bibliography{ref}

\newpage
\appendix
\onecolumngrid

\section{Leakage detection} \label{app:leakage_detection}

Let us demonstrate that the action of the circuit of \cref{fig:leakage_detection} performs as stated. Let us first suppose that the input state $\ket{\psi}$ is in the computational subspace. In this case the CZ gates across pairs are equivalent to CZs between the dual-spin qubits (see \cref{fig:st_gates}). Therefore, for a state
\begin{equation}
    \ket{\psi} = \alpha\ket{0} + \beta\ket{1},
\end{equation}
the action of the circuit is:
\begin{align*}
    \ket{\psi}\ket{S} &= (\alpha\ket{0} + \beta\ket{1})\ket{-} \\
    &\xrightarrow{CZ} \alpha \ket{0}\ket{-} + \beta \ket{1}\ket{+} \\
    &\xrightarrow{H_1H_2} \alpha \ket{+}\ket{1} + \beta \ket{-}\ket{0} \\
    &\xrightarrow{CZ} \alpha \ket{-}\ket{1} + \beta \ket{-}\ket{0} \\
    &= \ket{S}X\ket{\psi}
\end{align*}
Thus in this case the measurement always returns a singlet and the state is preserved (up to an X flip that is deterministic and can be stored in memory).

If the initial state is leaked however, we cannot use the reduced formalism above and instead work with all four spins. For a state
\begin{equation}
    \ket{\psi} = \alpha\ket{\uparrow\uparrow} + \beta\ket{\downarrow\downarrow},
\end{equation}
the action of the circuit is:
\begin{align*}
    \ket{\psi}\ket{S} &= (\alpha\ket{\uparrow\uparrow} + \beta\ket{\downarrow\downarrow})\ket{S} \\
    &\xrightarrow{CZ} \alpha \ket{\uparrow\uparrow}\ket{S} + \beta \ket{\downarrow\downarrow}\ket{T_0} \\
    &\xrightarrow{H_1H_2} \alpha \ket{\uparrow\uparrow}\ket{\uparrow\downarrow} + \beta \ket{\downarrow\downarrow}\ket{\downarrow\uparrow} \\
    &\xrightarrow{CZ} \alpha \ket{\uparrow\uparrow}\ket{\uparrow\downarrow} - \beta \ket{\downarrow\downarrow}\ket{\downarrow\uparrow} \\
    &= \alpha \ket{T_+}\ket{0} + \beta \ket{T_-} \ket{1}
\end{align*}
where in the third line we used the fact that the action of H on an even state is identity (see \cref{sec:physical_implementation_summary}). The measurement of the first pair would therefore necessarily show an even parity, and the output state will be projected back onto the computational subspace in an unknown state ($\ket{0}$ or $\ket{1}$).

\section{Hook errors} \label{app:hook_errors}

We here explain why the choice of the stabiliser circuits presented in \cref{fig:stabilisers} does not lead to distance-reducing hook errors.

\begin{figure}
    \centering
    \includegraphics[width=.8\linewidth]{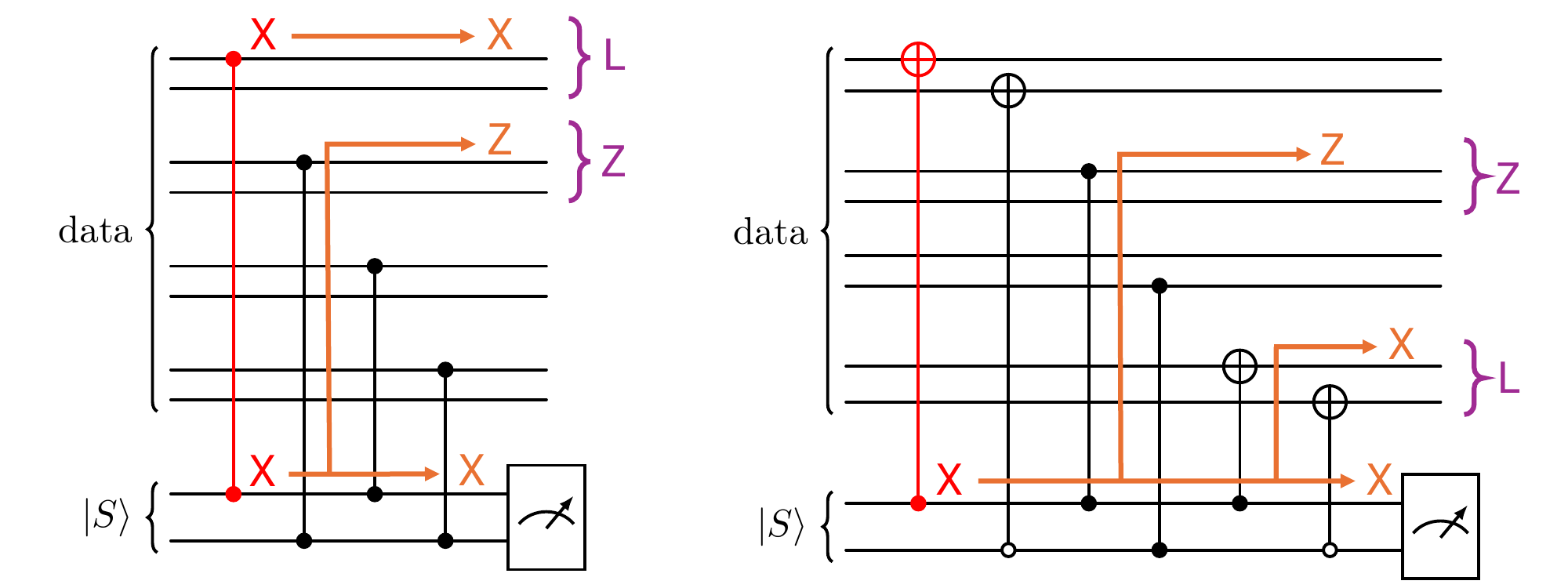}
    \caption{Hook error propagation in the implementation of a Z stabiliser (left) and an XZZX stabiliser (right). The first gate (in red) is erroneous and leads to one or two X spin-flips. Orange arrows show how errors propagate through the subsequent gates. Final errors patterns on the individual spins are showed in orange. How they translate at the singlet-triplet level for the data qubits is then written in purple (with L being a leaked state). In both cases, the hook error does not reduce the effective code distance as the leaked state is detectable.}
    \label{fig:hook_errors}
\end{figure}

Let us first examine the simplest case of a Z stabiliser, which only requires four CZs (\cref{fig:stabilisers} without the H gates). Error propagation can only occur if a given two-qubit gate is followed by an X or Y error on one of the spins of the ancilla, as this error would then anti-commute with the next gates. Since CZs here are exchange gates hence spin-conserving, they cannot flip the ancilla's spin without also flipping the data qubit's spin (see \cref{sec:physical_implementation_summary}, left). Thus, the only hook errors that can happen arise from an XX or a YY error on the first or second CZ, then propagating as a Pauli Z through the third or fourth CZ, as shown in \cref{fig:hook_errors}. This results in the first or second data qubit being leaked, and the third or fourth data qubit receiving a Z error. This is not a distance-reducing hook error per se, as leakage is detectable.

The same logic can be applied to X stabilisers. The only difference is the presence of a final round of Hadamard gates. Before the H's, the only possible error configurations arising from hook errors are a leaked data qubit and a phase flip on another data qubit (same as above). Since the Hadamards leave leaked states idle, the final error configuration would be leaked + X error. This would also preserve the distance as leakage is detectable.

In the XZZX scenario, the only difference lies in the fact that CNOTs do not necessarily create leakage in pairs. This is not problematic however, as shown in the right panel of \cref{fig:hook_errors}. The only opportunity for an error to propagate to multiple data qubits is again via an X or Y flip of a spin of the ancilla pair. Then propagation through subsequent gates can cause Pauli Z's or leakage. Two Pauli Z's cannot occur as the CZs are controlled by two distinct spins of the ancilla pair, and leakage is detectable. In this sense, there is no distance-reducing hook error in this case either.

\section{Noise models} \label{app:noise_models}

In this section, we explain why restricting ourselves to timing errors is a good representation of all the noise mechanisms at hand, and derive the form of the noise in each case \textit{i.e.} \cref{eq:fid_analytic_expressions1}, \cref{eq:fid_analytic_expressions2} and \cref{eq:fid_analytic_expressions3}. The assumption we make is that the noise stems from random fluctuations of the gate parameters with respect to the ideal ones due to \textit{e.g.} charge noise~\cite{jnane_ab_initio_2024}. This means that it must respect the structure of the gate's matrix, in particular for the exchange gate (which has a lot of zero terms).

\subsection{Driven single-qubit gate}

Let us first study a driven single-qubit gate. In the laboratory frame of reference, its matrix reads:
\begin{equation}
    U(t) = \begin{pmatrix}
        \left(\cos(\frac{\Omega_rt}{2})-i\frac{\delta \omega}{\Omega_r}\sin(\frac{\Omega_rt}{2})\right)e^{-i\frac{\omega t}{2}} & -\frac{i\Omega}{\Omega_r}\sin(\frac{\Omega_rt}{2})e^{-i\frac{\omega t}{2}}  \\
        -\frac{i\Omega}{\Omega_r}\sin(\frac{\Omega_rt}{2})e^{i\frac{\omega t}{2}} & \left(\cos(\frac{\Omega_rt}{2})+i\frac{\delta \omega}{\Omega_r}\sin(\frac{\Omega_rt}{2})\right) e^{i\frac{\omega t}{2}}
    \end{pmatrix}
\end{equation}
where $\Omega$ and $\omega$ are respectively the pulse amplitude and frequency, $\delta\omega$ is the difference between the qubit frequency and $\omega$ and
\begin{equation}
    \Omega_r = \sqrt{\Omega^2+\delta\omega^2}.
\end{equation}
Timing errors can be described as angular errors in the gate. While a perfect X gate requires $\Omega_r T=\pi$, an imperfect one may be characterised by $\Omega_r T=\pi+\delta\theta$. In this case, all elements of the matrix will be non-zero, meaning that the resulting noise will be quite general (it can be any Pauli X, Y or Z). Choosing to restrict ourselves to timing errors would only alter the weighting between these errors, which is not dramatic for our study.

Using the formula from \cref{eq:fid_formula}, one can extract the fidelity for a given set of imperfect parameters:
\begin{align*}
    \mathcal{F} &= \frac{2+\left|\frac{2\mathrm{i}\Omega}{\Omega_r}\cos(\delta\theta/2)\right|^2}{6} \\
        &\approx \frac{1}{3} + \frac{2}{3}\left(\frac{\Omega}{\Omega_r}\right)^2\left(1-\frac{\delta\theta^2}{4}\right)
\end{align*}
for small values of $\delta\theta$. Now integrating over a normal distribution of angles of width $\sigma_{\delta\theta}$, one obtains:
\begin{equation}
    \overline{\mathcal{F}} = \frac{1}{3} + \frac{2}{3}\left(\frac{\Omega}{\Omega_r}\right)^2 \left(1-\frac{\sigma_{\delta\theta}^2}{4}\right)
\end{equation}
When all parameters but the gate time are perfect, $\Omega_r = \Omega$. Additionally noting that $\sigma_{\delta\theta} = \Omega_r\sigma_T$, one obtains the expression of \cref{eq:fid_analytic_expressions1}.

\subsection{Exchange gates}

The general form of an exchange gate in the laboratory frame of reference is~\cite{meunier_cz}:
\begin{equation}
      U(t) = \begin{pmatrix}
        e^{-iE_Z t} & 0 & 0 & 0 \\
        0 & (\cos(\frac{\Omega_r t}{2})-i\frac{\Delta E_Z}{\Omega_r}\sin(\frac{\Omega_r t}{2}))e^{i\frac{Jt}{2}} & -i\frac{J}{\Omega_r}\sin(\frac{\Omega_r t}{2})e^{i\frac{Jt}{2}} & 0 \\
        0 & -i\frac{J}{\Omega_r}\sin(\frac{\Omega_r t}{2})e^{i\frac{Jt}{2}} &(\cos(\frac{\Omega_r t}{2})+i\frac{\Delta E_Z}{\Omega_r}\sin(\frac{\Omega_r t}{2}))e^{i\frac{Jt}{2}}  & 0 \\
        0 & 0 & 0 & e^{iE_Z t}
    \end{pmatrix}
\end{equation}
where $E_Z$ and $\Delta E_Z$ are respectively the average and the difference of the Zeeman splitting of the qubits, $J$ is the exchange coupling and
\begin{equation}
    \Omega_r = \sqrt{J^2+\Delta E_Z^2}.
\end{equation}
Contrary to the driven single-qubit gate, this matrix contains many zero entries (related to the spin-conservation property of the exchange gate). As explained many times in the main text this means that the noise has a specific structure (it cannot be described by a generic depolarising channel). When looking at timing errors in the form $\Omega_rT = \Omega_rT_0 + \delta\theta$, one can see that the matrix will take the most general form with six non-zero elements. Again, this justifies our focus on timing errors only.

\subsubsection{CZ gate}

The CZ gate is obtained by setting $J = \Delta E_Z/\sqrt{3}$ and $\Omega_rT_0 = \pi$, and by applying a single-qubit Z gate of angle $E_ZT_0/2$ on each qubit. Note that these Z gate can be applied virtually \textit{i.e.} in software~\cite{virtual_z_gates} therefore they do not contribute to the final infidelity. When all parameters except the gate time are perfect, one obtains:
\begin{align*}
    \mathcal{F} &= \frac{4 + \left|2 + 2\cos(\delta\theta/2)\right|^2}{20} \\
        &\approx 1 - \frac{\delta\theta^2}{10}
\end{align*}
for small $\delta\theta$. This leads to the expression given in \cref{eq:fid_analytic_expressions2} when integrating over a normal distribution of $\delta\theta$.

\subsubsection{H gate}

The Hadamard gate is obtained in a similar fashion, by setting $J = \Delta E_Z$ and $\Omega_rT_0 = \pi/2$, and again by applying a single-qubit Z gate of angle $E_ZT_0/2$ on each qubit. One can verify that the fidelity in this case takes the same form as for the CZ.

\subsection{CNOT gate}

We finally focus on the CNOT gate, implemented by targetting the $\ket{\uparrow\uparrow}\leftrightarrow\ket{\uparrow\downarrow}$ transition with a driving field. Let us first write the Hamiltonian of the corresponding operation in the frame rotating at the drive frequency in the regime $J\ll\Delta E_Z$ and near the resonant frequency~\cite{2q_gate_russ}:
\begin{align}
    H \approx \frac{1}{2}\begin{pmatrix}
        2(E_z-\omega) & (\Omega_2+\frac{J\Omega_1}{2\Delta E_z}) & 0 & 0 \\
        (\Omega_2+\frac{J\Omega_1}{2\Delta E_z}) & -J + (\Delta E_z+\frac{J^2}{2\Delta E_z}) & 0 & 0 \\
        0 & 0 & -J - (\Delta E_z+\frac{J^2}{2\Delta E_z}) & (\frac{J\Omega_1}{2\Delta E_z}-\Omega_2) \\
        0 & 0 & (\frac{J\Omega_1}{2\Delta E_z}-\Omega_2) & -2(E_z-\omega)
        \end{pmatrix}.
\end{align}
The Hamiltonian is block-diagonal, making the calculation of the time-evolution operator easy:
\begin{equation}
    U(t) = \begin{pmatrix}
        R_1(\theta_1(t)) & 0 \\
        0 & R_2(\theta_2(t)) \\
    \end{pmatrix}
\end{equation}
where $R_i(\theta_i(t))$ are single-qubit rotations that depend on the physical parameters (exchange coupling, amplitude and frequency of the drive etc.). By carefully choosing $\omega$, $R_2$ becomes a rotation around the $x$-axis while $R_1$ is more general. A synchronisation condition can however be implemented such that for all $t$, $\theta_1(t)=2\theta_2(t)$. This way, if the perfect gate time $T_0$ is chosen such that $\theta_2(t)=\pi$, then $R_1=I$ and $R_2=X$. This leads to the desired matrix for the CNOT.

If in turn the rotation angles are erroneous due to timing errors, such that $\theta_2 = \theta_1/2 = \pi+\delta\theta$, then $U$ takes its general block-diagonal form, and the resulting infidelity is:
\begin{align*}
    \mathcal{F} &= \frac{4 + |2\cos^2(\theta) + 2\cos^2(\theta/2)|^2}{20} \\
                &\approx 1 - \frac{\delta\theta^2}{2}
\end{align*}
Once again, integrating over a Gaussian distribution of timing errors leads to the results stated in \cref{eq:fid_analytic_expressions3}.

\section{Simulations details} \label{app:simulations_details}

\paragraph{Stabiliser and leakage detection circuits simulations.} As explained in the main text, we wish to precisely describe the propagation of coherent errors within the stabiliser and leakage detection circuits. For this purpose, we make use of the Choi-Jamiolkowski (CJ) isomorphism~\cite{choi_completely_1975, jamiolkowski_linear_1972}, which is a method enabling the decomposition of a quantum circuit $\mathcal{E}$ (given as a black box) in terms of its Kraus operators $K_i$:
\begin{equation}
    \mathcal{E} = \sum_i p_i K_i^\dagger.K_i
\end{equation}
with $p_i\in[0,1]$ and $\sum_ip_i=1$.
For a system involving $n$ data qubits, this protocol requires the simulation of a circuit acting on $2n$ qubits. The precise steps of the CJ isomorphism are the following:
\begin{itemize}
    \item Prepare $n$ Bell pairs \textit{e.g.} in the $(\ket{00}+\ket{11})/\sqrt{2}$ state;
    \item Apply the quantum circuit to be analysed to half of each Bell pair
    \item Reshape the resulting $2^{2n}$ state vector into a $2^n \times 2^n$ matrix. This is the channel's matrix $U$.
\end{itemize}
Mathematically, this reads:
\begin{align*}
    \ket{\Psi} &= \bigotimes_{i=1}^n (\ket{00}+\ket{11})/\sqrt{2} = \sum_x \ket{x}\ket{x} \\
               &\xrightarrow{\text{Apply circuit}} \sum_x U\ket{x} \ket{x} \\
               &\xrightarrow{\text{Reshape}} \sum_x U\ket{x} \bra{x} = U\\
\end{align*}
Thus this protocol indeed results in the computation of the matrix $U$ of the quantum circuit. While this describes a purely unitary channel, this can be extended to the use of more general noise channels and to the addition of measured out ancilla qubits. In this case, instead of outputting a single matrix, one would output multiple unitary matrices depending on \textit{e.g.} the measurement outcome, each associated with a different probability. These form the Kraus operators.

In this paper, we study circuits acting on at most four qubits pairs, \textit{i.e.} eight spins, plus one ancilla pair. This therefore requires the simulation of $6\times 2+2=18$ qubits. For such large systems, we use the QuEST toolset which is an efficient library for the exact simulation of quantum states~\cite{jones_quest_2019}.

\begin{figure}
    \centering
    \includegraphics[width=0.8\linewidth]{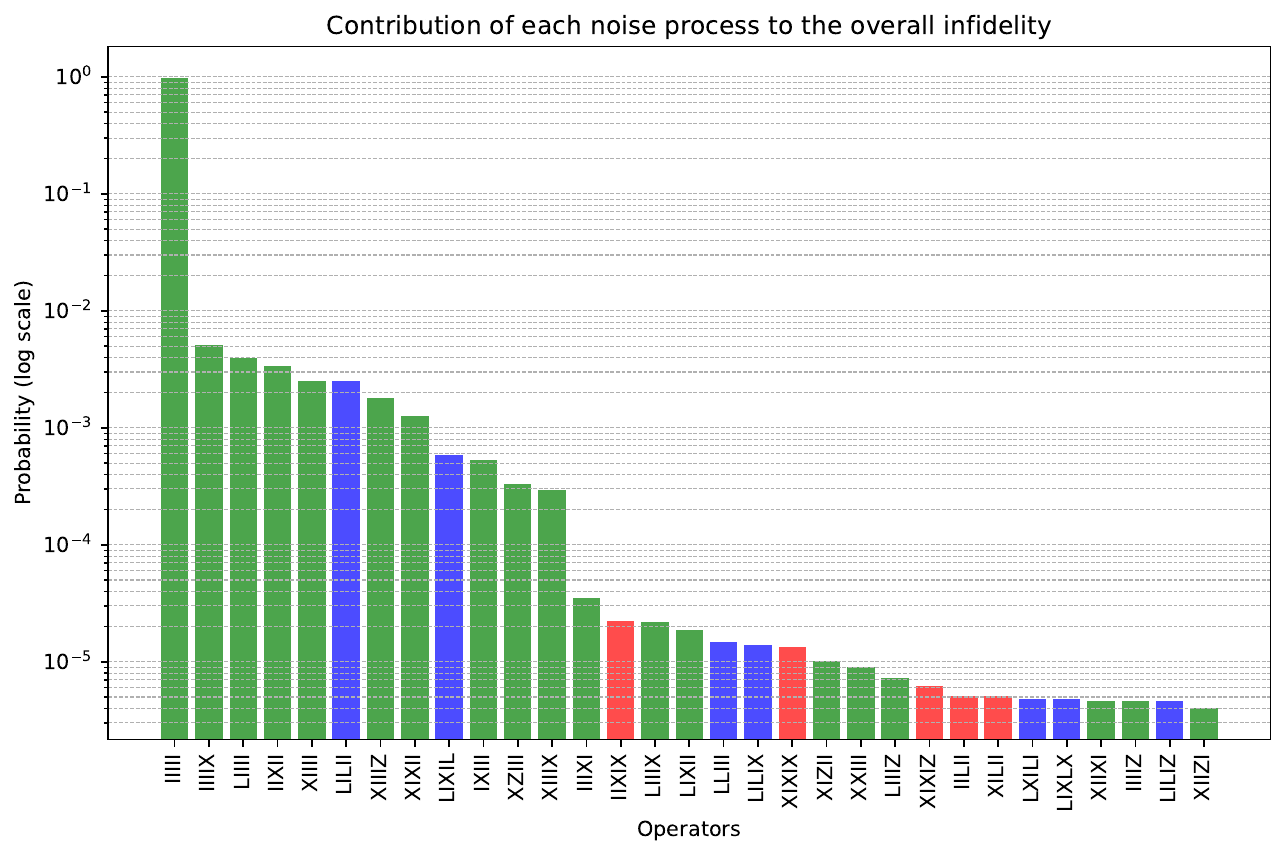}
    \caption{Distribution of post-twirling Pauli errors for an X stabiliser implemented with the circuit presented in \cref{fig:stabilisers}, at a physical error rate $p=0.6\%$. Green bars correspond to at most one data-qubit error and blue bars to pairs of leakage errors on an ancilla and a data qubit (with potentially another data qubit error). These phenomena cover all first-order events and are easily correctable. Red bars are second-order errors and more harmful to the code. These include errors on more than one data qubit or unpaired leakage events.}
    \label{fig:lookup_table_bar_chart}
\end{figure}

\paragraph{Twirling approximation.} At this point, the action of the stabiliser circuit is perfectly described and the Kraus operators $K_i$ are general (they can contain coherent errors). We now wish to twirl them such that only Pauli errors are fed to the decoder. Denoting $K_0$ the perfect action of the stabiliser circuit, $K_iK_0^{-1}$ only describes the noise and can be written in the Pauli basis as:
\begin{equation}
    K_iK_0^{-1} = \sum_j \alpha_{i,j}P_j
\end{equation}
where $\alpha_{i,j}$ is a complex coefficient. Note that this step is computationally intensive as we consider matrices of size $2^8\times 2^8$ (8 data spins per stabiliser). Applying the twirling approximation, every Kraus operator is replaced with a twirled version $\tilde{K}_i$~\cite{Cai_2019}
\begin{equation}
    \tilde{K}_iK_0^{-1} = \sum_j |\alpha_{i,j}|^2P_j.
\end{equation}
With this transformation one can now write
\begin{align}
    \mathcal{E}\circ\mathcal{E}_0^{-1} = \sum_j \left( \sum_i p_i|\alpha_{i,j}|^2 \right) P_j^\dagger.P_j
\end{align}
This shows that the new Kraus operators are the Pauli group, with associated weights given in the brackets. For every circuit considered in this paper (stabiliser or leakage detection), we can for instance plot the relative importance of each Pauli mechanism, as represented in \cref{fig:lookup_table_bar_chart}. Note that this is also dependent on the physical parameters such as the exchange coupling or the $g$-factors.

\paragraph{QEC simulator.} The previous step gave us the Pauli errors that are associated with the application of a given stabiliser or leakage detection circuit. We now feed that into a QEC simulator that:
\begin{itemize}
    \item Introduces random Pauli errors after the application of every stabiliser and leakage detection circuit;
    \item Computes the syndrome according to these Paulis (accounting for the modified action of the Hadamard gate in case of leakage);
    \item Decodes the syndrome using MWPM and the leakage information;
    \item Checks for a logical error by adding the correction to the existing errors
\end{itemize}
Note that that we make the conventional assumption that the data qubits are measured out at the final round. This is equivalent to assuming a perfect final round of stabiliser measurements and leakage detections.

\paragraph{Monte-Carlo simulations.} The logical error rates are then computed by repeating this experiment many times (between 10,000 and 10,000,000 depending on the logical error rates).

\end{document}